\documentstyle[12pt,agums,psfig]{article}

\lefthead{AFRAIMOVICH ET AL.}
\righthead{GPS TESTING OF THE TRANSIONOSPHERIC RADIOCHANNEL}
\received{}
\revised{}
\accepted{}
\paperid{}

\authoraddr{E.~L. Afraimovich, V. A. Karachenschev,
Institute of Solar-Terrestrial Physics SD RAS,
p.~o.~box~4026, Irkutsk, 664033, Russia,
fax: +7 3952 462557; e-mail:~afra@iszf.irk.ru}

\begin{document}

\title{Testing of the transionospheric radiochannel
using data from the global GPS network}

\author{Afraimovich~E.~L., Karachenschev~V.~A.\\
Institute of Solar-Terrestrial Physics SD RAS,\\
p.~o.~box~4026, Irkutsk, 664033, Russia\\
fax: +7 3952 462557; e-mail:~afra@iszf.irk.ru}

\date{}

\begin{abstract}

Using the international ground-based network of two-frequency
receivers of the GPS navigation system provides a means of
carrying out a global, continuous and fully-computerized
monitoring of phase fluctuations of signals from satellite-borne
radio engineering systems caused by the Earth's inhomogeneous and
nonstationary ionosphere. We found that during major geomagnetic
storms, the errors of determination of the range, frequency
Doppler shift and angles of arrival of transionospheric radio
signals exceeds the one for magnetically quiet days by one order
of magnitude as a minimum. This can be the cause of performance
degradation of current satellite radio engineering navigation,
communication and radar systems as well as of superlong-baseline
radio interferometry systems.
\end{abstract}

\begin{article}

\section{Introduction}
\label{SPE-sect-1}
Radio engineering satellite systems (RESS), with their
ground-based and space-borne support facilities, are finding
ever-widening application in various spheres of human activity.
They are able to provide global coverage, accuracy, continuity,
high reliability and meet a number of other requirements imposed
when tackling a broad spectrum of engineering problems. However,
the use of RESS also implies new (and, in some cases, more
stringent) requirements dictated by the need to ensure safety and
economical efficiency of the operation of ground-based and
airborne facilities, as well as to solve special problems
(observation, aerophotography, searching and rescue of distressed
transport vehicles and people). This applies equally for
performance of Global Navigation Satellite Systems (GNSS) as well
as for very-long-baseline radio interferometers (VLBI)
[{\it Thompson,~R. et al}., 1986].

Such requirements cannot be met unless the influence of
destabilizing factors in the radio wave propagation channel is
taken into account. As electromagnetic waves propagate through
the ionosphere, they experience quite varied disturbances
[{\it Davies}, 1969; {\it Kolosov et al}., 1969;
{\it Yakovlev}, 1985;
{\it Goodman and Aarons}, 1990; {\it Yakubov}, 1997].

The key characteristic of the ionosphere that determines the
variation of radio wave parameters is the integral (total)
electron content (TEC) $I(t)$ or its derivatives (with respect to
time and space) $I^{'}_t$, $I^{'}_x$ and $I^{'}_y$ along the
propagation path.

TEC variations may be arbitrarily classified as regular and
irregular. Regular changes (seasonal, diurnal) - for the
magnetically quiet mid-latitude ionosphere at least - are
described by models providing relative accuracy of TEC prediction
in the range 50...80$\%$. Irregular changes (variations) are
associated with ionospheric irregularities of a different nature,
the spectrum of which has a power law character 
[{\it Gajlit}, 1983; {\it Gershman et al}., 1984; 
{\it Yakubov}, 1997].

TEC variations introduce proportionate changes of the signal
phase $\varphi(t,x,y)=k_1I(t,x,y)$, which gives rise to measuring
errors of the range $\sigma{D}=k_2dI$,
the frequency Doppler shift of the signal
$\sigma{f}=k_3I^{'}_t$, and the
angles of arrival of the radio wave $\sigma\alpha_x=k_4I^{'}_x$
and $\sigma\alpha_y=k_4I^{'}_y$, because the last
four quantities are proportional to the time and space
derivatives of the phase. Furthermore, the maximum value of the
measuring error of angular deviations can be deduced using the
relation $\sigma\alpha=k_4~\sqrt{(I^{'}_x)^{2}+(I^{'}_y)^{2}}$.

The coefficients $k_1-k_4$ are inversely proportional to the signal
frequency $f_c$ or to its square 
[{\it Davies}, 1969; {\it Kravtsov et al}., 1983;
{\it Goodman and Aarons}, 1990].
A calculation uses a Cartesian topocentric coordinate
system with the axis $x$ pointing eastward $E$, and the axis $y$
pointing northward $N$.

Investigations of phase fluctuations of transionospheric signals
have been and are carried out using radio beacons on satellites
with circular and geostationary orbits 
[{\it Komrakov and Skrebkova}, 1980; {\it Livingston et al}., 
1981; {\it Gajlit}, 1983]. 
The trouble with
these measurements is that temporal and spatial resolution is
low, and continuity and global coverage of observations are
unavailable.

The use of the international ground-based network of
two-frequency receivers of the navigation GPS system that at the
beginning of 2002 consisted of no less than 1000 sites and is
posting its data on the Internet, opens up new avenues for a
global, continuous, fully computerized monitoring of phase
fluctuations of signals and associated errors of RESS
performance.

Some research results on the prediction and estimation of radio
signal fluctuations and errors of RESS performance caused by them
were reported in earlier work
[{\it Afraimovich and Karachenschev}, 2002].

Results of experimental investigations of phase fluctuations of
RESS signals that are discussed in this paper were derived from
analyzing the data for a set consisting of 100 to 300 GPS
stations covering the time interval 1998-2001, for different
conditions of geomagnetic activity (the $D_{st}$-index from -13 to
-377 $nT$). The number of passes of GPS satellites with a duration
of no less than 2.5 hours, the data from which were used in the
analysis, totaled no less than 300000, or several orders of
magnitude in excess of the currently available statistic of such
measurements.

Below we give an outline of the techniques used in this study and
illustrate their application in the analysis of ionospheric
effects. The overall sample statistic of errors $\sigma{D}$,
$\sigma{f}$ and $\sigma\alpha$
is presented for different geomagnetic conditions. To ease
comparison with other research results reported in
[{\it Gajlit}, 1983; {\it Kravtsov et al}., 1983], the
errors $\sigma{D}$, $\sigma{f}$ and $\sigma\alpha$ are
calculated for the working frequency
of 300 MHz.

\section{General information about the database used}
\label{SPE-sect-2}
This study relies on the data from the global network of
receiving GPS stations available on the Internet (Fig. 1). As is
evident from Fig. 1, the receiving sites are relatively dense on
the territory of North America and Europe, and less as dense in
Asia. Fewer stations are located on the Pacific and Atlantic.

Such coverage of the terrestrial surface by GPS receivers makes
it possible, already at the present time, to address the problem
of a global investigation of ionospheric disturbances and their
consequences with a very large spatial accumulation.

Thus, in the Western hemisphere the corresponding number of
stations is as large as 500, and the number of 
line-of-sight (LOS) to the
satellite is at least 2000...3000.
This provides a number of
statistically independent series at least two orders of magnitude
higher than would be realized by recording VHF radio signals from
first-generation geostationary satellites or low-orbit navigation
satellites - TRANSIT [{\it Gershman et al}., 1984].

Our analysis used the North-American sector with a large number
of GPS station -- range of longitudes -120$^\circ$ ...
-60$^\circ$ $E$ and of latitudes 20$^\circ$ ...
70$^\circ$ $N$. Table~1 presents information about number of
the days, the number $n$ of GPS arrays composed of three
stations, the data from which are used, and extreme
values of $D_{st_{min}}$. Total number of spectra $\Sigma$ also
is resulted which were used for definition of average sizes of
inclinations.

For a variety of reasons, for the various events to be analyzed,
we selected somewhat differing sets of receiving stations, yet
the geometry of the experiment was virtually identical for all
events. The coordinates of the receiving stations used in the
experiment were available at the electronic address:
ftp://lox.ucsd.edu/pub.processing/gamit/setup/coords.txt.

\section{Analysis of the measuring errors of the range, Doppler
frequency and angles of arrival of the radio wave caused by
changes in the regular ionosphere}
\label{SPE-sect-3}

Recently a number of authors [{\it Wilson et al}., 1995;
{\it Mannucci et
 al}., 1998; {\it Schaer et al}., 1998; {\it and others}] have
developed a new
technology for constructing Global Ionospheric Maps (GIM) of 
TEC using IONEX data from the international IGS-GPS network.
The GIM technology and its uses have been reported in a large
number of publications [{\it Wilson et al}., 1995;
{\it Mannucci et al}., 1998].

The standard IONEX format is
described in detail in [{\it Schaer et al}., 1998].
Therefore, we will
not give a detailed description of the GIM technology for
reasons of space but limit ourselves only to the information
required for the presentation of our method.
Two-hour TEC maps are easily accessible to any user, which are
calculated by several research groups in the USA and Europe and
are availabe on the Internet in the standard IONEX format
(ftp://cddisa.gsfc.nasa.gov/pub/gps/products/ionex). It is
also possible to obtain 15-min maps if necessary.

Fig.~2 is a schematic representation of a single elementary GIM
cell. The cell nodes are designated as $a$, $b$, $c$, $d$. The
cell size ($5^\circ$ in longitude and $2.5^\circ$ in latitude) is
determined by the IONEX file standard. For simplifying the
transformations to an approximation sufficient for our problem
for latitudes not exceeding $60^\circ$, the cell can be
represented as a rectangle with the sides $d_e$ and $d_n$. It is
easy to overcome this limitation by complicating to a certain
extent the transformations allowing for the sphericity; however,
we do not present them in this report.

The linear size of the rectangular cell in latitude is
independent of the latitude and is $d_e=279$ km; the linear size
in longitude depends on the latitude, and for $40^\circ N$ it is
$d_n=436$ km.

For each time $t$, for the nodes $a$, $b$, $c$, $d$
from the IONEX file the values of vertical TEC are known -
$I_a$, $I_b$, $I_c$, $I_d$.

The determination of the range $D$ by the phase method is based on
measuring the phase difference $\varphi$ between the
received signal and the reference signal formed in the receiver.
Such a measurement can be made at the intermediate or carrier
frequency of the signal. In this case:

\begin{equation}
\label{EQ-eq-01}
D=c \times \frac{\varphi}{2~\pi~f_c}
\end{equation}

where $c$ is the propagation velocity of radio waves in a
free space.

Generally the quantity $\varphi$ for the transionospheric
propagation may be regarded as the sum of two components
[{\it Afraimovich et al}., 1998]:

\begin{equation}
\label{EQ-eq-02}
\varphi=\varphi_s{+}\Delta\varphi
\end{equation}

where $\varphi_s$ is the main component associated with a change
of the distance between the signal source and the receiver.

Analysis of the measuring errors of the range, Doppler
frequency and angles of arrival of the radio wave caused by
changes in the regular ionosphere investigations of global phase
variations of radio signals and their influence on the operation
of RESS ought to take into account the proportionate relationship
between phase (phase derivative) changes of the transionospheric
signal and corresponding TEC variations 
[{\it Kravtsov et al}., 1983; {\it Goodman and Aarons}, 1990]:

\begin{equation}
\label{EQ-eq-03}
\Delta\varphi=8.44 \times 10^{-7} \times\frac{I_{a(b,c,d)}}{f_c}+\varphi_0
\end{equation}

where $f_c$ is the radio wave frequency (Hz); $I_{a(b,c,d)}$ is
the TEC measured at the points $a$, $b$, $c$, and $d$
($10^{16}~^{el}/_{m^2}$); and
$\varphi_0$ is the initial phase
[{\it Spoelstra and Kelder}, 1984].

By way of example we now analyze the errors of measurement of the
range $\sigma{D}$, the frequency Doppler shift $\sigma{f}$ and
the angle of arrival of the radio wave $\sigma\alpha$ using
the IONEX data and the method that was developed at the ISTP
SB RAS [{\it Afraimovich and Karachenschev}, 2002].

The change of the range that is introduced by the ionosphere
(ionospheric error) is proportional to $\Delta\varphi$:

\begin{equation}
\label{EQ-eq-04}
\sigma{D}=c \times \frac{\Delta\varphi}{2~\pi~f_c}
\end{equation}

Upon substituting (3) into (4), we can obtain the expression for
determining the measuring error of the range $\sigma{D}$
introduced by the ionosphere:

\begin{equation}
\label{EQ-eq-05}
\sigma{D}=\frac{c \times 8.44 \times 10^{-7} \times dI}
{2~\pi~f^{2}_c}=4.48 \times dI
\end{equation}

As is seen from (5), the error $\sigma{D}$ is directly
proportional to the TEC variation $dI$ and inversely
proportional to the carrier frequency squared.

Using the values of the spatial derivatives of TEC $I^{'}_x$
and $I{'}_y$
and of the derivative of TEC with respect to time $I^{'}_t$
makes it
possible to uniquely obtain - for each instant of time - the
values of errors of determination of the angle of arrival
$\sigma\alpha$ and
the frequency Doppler shift $\sigma{f}$ by formulas
[{\it Kravtsov et al}., 1983; {\it Goodman and Aarons}, 1990]:

\begin{equation}
\label{EQ-eq-06}
\sigma{f}=\frac{1.39 \times 10^{2}}{f^{2}_c}
\times \sqrt{(I^{'}_x)^{2}+(I^{'}_y)^{2}}
\end{equation}

\begin{equation}
\label{EQ-eq-07}
\sigma\alpha=\frac{1.34 \times 10^{-7}}{f_c} \times I^{'}_t
\end{equation}

In the simplest case, the values of the derivatives for the
selected cell of the map can be obtained using TEC increments
for the four cell nodes and for two times $t_1$ and $t_2=t_1+d_t$:

\begin{equation}
\label{EQ-eq-08}
\begin{array}{rl}
dI=(I_{a2}-I_{a1}+I_{b2}-I_{b1}+I_{c2}-I_{c1}+I_{d2}-I_{d1})/4\\
\Delta I^{'}_t=(I_{a2}-I_{a1}+I_{b2}-I_{b1}+I_{c2}-I_{c1}+I_{d2}-I_{d1})/4d_t\\
\Delta I^{'}_x=(I_{c1}-I_{b1}+I_{d1}-I_{a1}+I_{c2}-I_{b2}+I_{d2}-I_{a2})/4d_e\\
\Delta I^{'}_y=(I_{a1}-I_{b1}+I_{d1}-I_{c1}+I_{a2}-I_{b2}+I_{d2}-I_{c2})/4d_n
\end{array}
\end{equation}

Where necessary, the spatial derivatives can be estimated
by taking into account the TEC values in adjacent nodes of the
map, and the time derivative (with a time resolution of IONEX
maps no worse than 15 min) can be inferred by averaging
increments for several successive time counts.

The procedures (5), (6) and (7) are performed for all cells
of the
selected spatial range and for the selected time interval of the
day. One variant of data representation implies a full
exploitation of the IONEX format with the difference that,
rather than the
values of TEC and the error of TEC determination [{\it Schaer
et al}., 1998],
are entered into the corresponding cells of the map values
of error $\sigma{D}$, $\sigma{f}$ and $\sigma\alpha$.

By way of example, it is appropriate to give the results derived
from analyzing the regular part of the spatial-temporal TEC
variations for a relatively magnetically quiet day of July 29,
1999 (with the largest deviation of the $D_{st}$-index of -40 $nT$)
and
for the magnetically disturbed day of April 6, 2000 (with the
largest deviation of the $D_{st}$-index of -293 $nT$).

Fig.~3 $a$, $b$, $c$ portrays the maps of the errors
$\sigma{D}$, $\sigma{f}$ and $\sigma\alpha$
obtained on the basis of files in the IONEX format for the
magnetically quiet day of July 29, 1999 in the geographic
coordinate system in the range of longitudes
-120$^\circ$ ... -60$^\circ$ $E$ and
latitudes 20$^\circ$ ... 70$^\circ$ $N$.
Fig.~3 $d$, $e$, $f$, respectively,
characterizes the values of $\sigma{D}$, $\sigma{f}$ and
$\sigma\alpha$ for the magnetically
disturbed day of April 6, 2000. The figure also shows the time
interval 19-21 UT, for which the analysis was carried out.
Contours show the values of errors of phase measurements in
units, respectively, of $\sigma{D}$ - "m" (meters),
$\sigma{f}$ - "Hz" (Hertz), and $\sigma\alpha$
- "arcmin" (minutes of arc). The vertical calibrated scale shows
the maximum and minimum values of the corresponding errors.

The above maps are a pictorial rendition of the behavior dynamics
of the errors $\sigma{D}$, $\sigma{f}$ and
$\sigma\alpha$ in the spatial and temporal ranges
selected. Noteworthy is a considerable difference of the maps for
the magnetically quiet and disturbed days. As is evident from
Fig.~3, gradients of spatial distribution of the errors during
disturbances increase more than an order of magnitude when
compared to the quiet period, which would lead to a degradation
of RESS performance.

\section{Analysis of the irregular errors $\sigma{D}$, $\sigma{f}$
and $\sigma\alpha$}
\label{SPE-sect-4}

Our analysis of the irregular component of the errors was based
on using raw data in the form of series of TEC values for
selected receiving sites as well as values of elevations
$\Theta_s{(t)}$ and
azimuths $\alpha_s{(t)}$ to visible satellites that were calculated
using a
specially developed program, CONVTEC, to convert RINEX-files
(standard files for the GPS system) available from the Internet
[{\it Afraimovich et al}., 1998].

Fig.~4 presents the geometry of transionospheric radio sounding.
The axes $z,y,x$ are pointing, respectively, to the zenith, the
north $N$, and to the east $E$; $P$ is the point of intersection
of the
LOS to the satellite (the line connecting the satellite to the
radio signal receiver) with the ionospheric $F_2$ region peak;
$S$ is
the subionospheric point (projection of the point $P$ onto the
terrestrial surface).

The GPS technology provides the means of estimating TEC
variations on the basis of phase measurements of TEC $I$ in each
of the spaced two-frequency GPS receivers using the formula
[{\it Hofmann-Wellenhof et al}., 1992]:

\begin{equation}
\label{EQ-eq-09}
I_0=\frac{1}{40{.}308}\frac{f^2_1f^2_2}{f^2_1-f^2_2}
[(L_1\lambda_1-L_2\lambda_2)+{\rm const}+nL],
\end{equation}

where $L_1\lambda_1$ and $L_2\lambda_2$~ are phase path
increments of the radio signal, caused by the phase delay in the
ionosphere (m); $L_1$ and $L_2$~ are the number of full phase
rotations, and $\lambda_1$ and $\lambda_2$ are the wavelengths
(m) for the frequencies $f_1$ and $f_2$, respectively; $const$~
is some unknown initial phase path (m); and $nL$~ is the error in
determination of the phase path (m).

Series of the values of elevations $\Theta_s(t)$ and azimuths
$\alpha_s(t)$ of the beam to the satellite were used to determine
the coordinates of subionospheric points, and to convert the
"oblique" TEC $I_{0}(t)$ to the corresponding value of the
"vertical" TEC $I(t)$ by employing the technique reported by
[{\it Klobuchar}, 1986]:

\begin{equation}
\label{EQ-eq-10} I = I_0 \times cos
\left[arcsin\left(\frac{R_z}{R_z + h_{max}}cos\Theta_s\right)
\right],
\end{equation}

where $R_{z}$ is the Earth's radius, and $h_{max}$=300 km is the
height of the $F_{2}$-layer maximum.

All results in this study
were obtained for elevations $\Theta_s(t)$ larger than
30$^\circ$.

To analyze the errors $\sigma{D}$, $\sigma{f}$ and $\sigma\alpha$
that are cased by the
irregular component of TEC variation, we make use of the
relations for the respective regular errors (5), (6) and (7).
The difference in the analysis of irregular errors implies a
different (compared with the IONEX technique) method of
determining the values of TEC and its derivatives
[{\it Afraimovich and Karachenschev}, 2002].

The phase differences $\Delta\varphi_{x,y}$ along the axes
$x$ and $y$ are
proportional to the values of the horizontal components of TEC
gradient $G_E=I^{'}_x$ and $G_E=I^{'}_y$.

To calculate the components of the
TEC gradient $I{'}_x$ and $I{'}_y$ are used
linear transformations of the differences of the values of the
filtered TEC $(I_B-I_A)$ and $(I_B-I_C)$ at the receiving
points $A$, $B$ and $C$:

\begin{equation}
\label{EQ-eq-11}
I{'}_x=\frac{y_A(I_B-I_C)-y_C(I_B-I_A)}{x_A y_C - x_C y_A};~~~
I{'}_y=\frac{x_C(I_B-I_A)-x_A(I_B-I_C)}{x_A y_C - x_C y_A}
\end{equation}

where $x_A$, $y_A$, $x_C$, $y_C$ are the coordinates of the sites
$A$ and $C$ in the topocentric coordinate system. When deriving
(11) we took into account that $x_B=y_B=0$, since site $B$ is the
center of topocentric coordinate system.

The time derivative of TEC $I{'}_t$ is determined by
differentiating $I(t)$ at the point $B$.

The procedures of (5), (6) and (7) are performed for all groups
of three GPS stations of the selected spatial range and for
satellites visible from these stations, as well as for the
selected time interval of the day.

Fig.~5 presents the results derived from analyzing the irregular
component of the errors $\sigma{D}$, $\sigma{f}$ and
$\sigma\alpha$ in the form of fluctuation
spectra of the range, Doppler frequency and angles of arrival of
radio waves.

With the purpose of improving the statistical reliability
of the data, we used the spatial averaging technique for
spectra within the framework of a novel technology
[{\it Afraimovich et al}., 2001].
The method implies using an appropriate
processing of TEC variations that are determined from the GPS
data, simultaneously for the entire set of GPS satellites (as
many as 5--10 satellites) "visible" during a given time interval,
at all stations of the global GPS network used in the analysis.

Individual spectra of the errors $\sigma{D}$, $\sigma{f}$ and
$\sigma\alpha$ were obtained by
processing continuous series of $I(t)$ measurements of a duration
no shorter than 2.5 hours. To eliminate errors caused by the
regular ionosphere, as well as trends introduced the motion of
satellites, we used the procedure of removing the linear trend by
preliminarily smoothing the initial series with the selected time
window of a duration of about 60 min.

Fig.~5 shows the overall character of the TEC variations $dI(t)$
that were filtered from TEC series obtained from measurements of
the phase difference between two coherently coupled signals from
the GPS system [{\it Hofmann-Wellenhof}, 1992] for the magnetically
quiet
day of July 29, 1999 (panel $a$, at the left) and a major magnetic
storm of July 15, 2000 (panel $e$, at the right). Furthermore, the
panels show the station names and locations, as well as GPS
satellite numbers (PRN).

As is evident from the figure, the intensity $dI(t)$ during the
disturbance even at such low latitudes is increased an order of
magnitude as a minimum. This is reflected on logarithmic
amplitude spectra $lgS(F)$ of TEC fluctuations and their
derivatives (left-hand scale in the figures) and of the
fluctuations of $\sigma{D}$, $\sigma{f}$ and $\sigma\alpha$,
converted to the working frequency
of 300 $MHz$ (right-hand scale) which are represented on a
logarithmic scale (panels $b$, $c$, $d$, $f$, $g$, $h$).

The logarithmic amplitude spectrum $lg S(F)$
obtained by using a standard FFT procedure.
Incoherent summation of the partial
 mplitude spectra $lg S(F)_i$
of different LOS was performed by the formula:

\begin{equation}
\label{EQ-eq-12} lg \langle S(F) \rangle =
lg \left[\frac{\sum\limits^n_{i=1}S(F)_i}{n}\right]
\end{equation}

where $i$ is the number of LOS; $i=$ 1, 2, ... $n$.

As a consequence of the statistical independence of partial
spectra, the signal/noise ratio, when the average spectrum is
calculated, increases due to incoherent accumulation at least by
a factor of $\sqrt{n}$, where $n$ is the number of LOS.

Fluctuation spectra from the magnetically quiet day of July 29,
1999 are shown by the thin line in panels $f$, $g$, $h$ (Fig.~5) 
for
comparison with the spectra from the disturbed day. The range of
fluctuation periods is shown in bold type along the abscissa axis
in panels $d$ and $h$. Panels $b$ and $f$ show also the number $n$
of GPS arrays composed of three stations, the data from which are
used to estimate the spatial derivatives of TEC
[{\it Afraimovich et al}., 2001].

Spectra of phase fluctuations have a power law character with the
values of the slopes $\nu$, shown in
panels $b$,
$c$, $d$, $f$, $g$, $h$. The slope of spectrum is determined by the
slope of the fitted straight line (thin black line in panel $b$
of Fig.~5). These results are in reasonably good agreement with
data reported in 
[{\it Komrakov and Skrebkova}, 1980; {\it Livingston et al}., 1981;
{\it Gajlit et al}., 1983; {\it Kravtsov et al}., 1983; 
{\it Gershman et al}., 1984;
{\it Yakubov}, 1997].

\section{Distribution of slopes and scales of spectra of the errors}
\label{SPE-sect-5}
In the preceding section we obtained the fluctuation spectra of
the errors $\sigma{D}$, $\sigma{f}$ and $\sigma\alpha$
corresponding to different ionospheric
conditions. It was shown that the spectra have a power law
character, with definite indices of the slope $\nu$ (Fig.~5).
However, the individual spectra that were obtained do not give a
full insight into the global picture of fluctuations of the
errors. To obtain a generalized estimate we carried out an
analysis of the spectra of errors obtained by processing the data
covering more than 30 days (over 600 spectra).

In our experiment the range of fluctuation periods of spectra
varies from 2 min to 2 hours. Experimental spectra of the errors
$\sigma{D}$, $\sigma{f}$ and $\sigma\alpha$
have a complicated form (Fig.~5). There are
spectral density maxima and minima. Generally, however, the
spectral density $S(F)$ decreases with the increasing fluctuation
frequency. For each individual spectrum, calculated on a
logarithmic scale, we determined the fitted straight line.

The result of a generalized estimation is represented by the
plots in Fig.~6. The plots show the distributions of the slopes
$P_{\sigma{D}}(\nu)$, $P_{\sigma{f}}(\nu)$,
$P_{\sigma\alpha}(\nu)$
of power law spectra of the errors $\sigma{D}$, $\sigma{f}$ and
$\sigma\alpha$,
respectively. The interval $\Delta\nu$ of the slopes of the
spectra $\sigma{D}$
is $\Delta_{\sigma{D}} = -1.45...2.45$, with the mean value
of ${<}\Delta_{\sigma{D}}> = -1.95$. In view
of the relationship between the spectrum
$\sigma{D} \sim \varphi{(t)}$
and the spectra $\sigma{f}$
and $\sigma\alpha$ that are proportional to the derivative
of $\varphi{(t,x,y)}$ ${(}\sigma{f} \sim \varphi^{'}(t)$ and
$\sigma\alpha \sim \varphi^{'}(x,y))$,
for the slopes of the spectra $\sigma{f}$ and $\sigma\alpha$
these values are,
respectively, $\Delta_{\sigma{f}} = -0.40...1.60$,
${<}\Delta_{\sigma{f}}> = -0.99$ and $\Delta_{\sigma\alpha}
= -0.40...2.00$,
${<}\Delta_{\sigma\alpha}> = -1.8$. Our results are in
good agreement with findings
reported by a number of authors [{\it Gajlit et al}., 1983].

The fitted straight line (the slant thin black line in panel $b$,
Fig.~5) for the logarithmic scale of errors in range may be
described by the expression (13):

\begin{equation}
\label{EQ-eq-13}
Y=a \times X+b
\end{equation}

here $a$ characterizes the slope of the straight line, and $b$
is the scale coefficient characterizing the rise of the straight
line with respect to the abscissa axis, that is, the value of
$Y_0$
when $X_0 = 0$.

The value of ${lgS}^{\sigma{D}}(F)$ can be determined from a single
(previously determined) set of values of $lg(F)$, $\nu_{\sigma{D}}$
and $b_{\sigma{D}}$ on
the basis of the expression:

\begin{equation}
\label{EQ-eq-14}
lgS^{\sigma{D}}(F)={\nu_{\sigma{D}}} \times lg(F)+b_{\sigma{D}}
\end{equation}

Using (14) and the resulting
value of $b_{\sigma{D}}$ and $\nu_{\sigma{D}}$
we can determine the value of any spectral component
of $lgS^{\sigma{D}}(F)$ and determine its contribution to the total
error.
Proceeding in a similar way, we can describe the spectral
characteristics of $\sigma{f}$ and $\sigma\alpha$.

Panels $a$, $b$, $c$ in Fig.~7 presents the distributions of the
values of the scales of power law spectra of the errors
${b}_{\sigma{D}}$, ${b}_{\sigma{f}}$ and ${b}_{\sigma\alpha}$
obtained for the entire set of data (over 600 spectra).

Let us estimate the amplitude $lgS^{\sigma{D}}_0$ of
the 32-minute harmonic $F_0$ of
the spectrum of the errors $\sigma{D}$ (Fig.~5) by making
use of the
values of ${<}\nu_{\sigma{D}}> = -1.96$ (Fig.~6) and
${<b}_{\sigma{D}}> = -6.59$ (Fig.~7). Upon
substituting the corresponding values into (14),
we obtain $lgS^{\sigma{D}}_0(F_0) = -0.153$,
which corresponds to $\sigma{D} = 0.70$ [m].

\section{Conclusion}
\label{SPE-sect-6}
Main results of this study are as follows:
\begin{enumerate}

\item
In this paper we have suggested a new technique for estimating
the errors of RESS performance, based on using phase measurements
from two-frequency receivers of the satellite navigation GPS
system.

\item
Fluctuation spectra of the errors are generally quite well
approximated by the power function. Mean values of the slopes of
spectra, obtained from the experiment, are in good agreement with
data reported by other authors.

\item
The errors $\sigma{D}$, $\sigma{f}$ and $\sigma\alpha$
are conveniently estimates with a
sufficient accuracy for practical purposes by using mean
statistical data for indices of slopes of spectra $\nu$
and scale coefficients $b$.

\item
Our results can be equally assigned to the estimation of the
Russian GNSS -- GLONASS performance quality.
\end{enumerate}

\acknowledgments
The authors are grateful to A.~A.~Neudakin for their
encouraging interest in this study and active participation in
discussions. The authors are also indebted to ~E.~A.~Kosogorov and
~O.~S.~Lesuta for preparing the input data. Thanks are also due
V.~G.~Mikhalkovsky for his assistance in preparing the English
version of the \TeX manuscript. This work was done with support
from the Russian Foundation for Basic Research (grants
01-05-65374 and 00-05-72026) and from RFBR grant of leading
scientific schools of the Russian Federation 00-15-98509.

{}
\end{article}

\newpage

\begin{figure}
\caption
{The map showing the locations of the
receiving stations of the global GPS system. The number of
receiving sites totals no less than 1000 as of the beginning of
2002.}
\label{fig1}
\end{figure}

\begin{figure}
\caption
{Geometry of an elementary GIM cell.
Nodes of the cell are denoted $a$, $b$, $c$, and $d$.
The cell size is determined by the values of
${d}_n$ = 279 km and of ${d}_e$ dependent on
the cell's location latitude.}
\label{fig2}
\end{figure}

\begin{figure}
\caption
{Maps of global distribution of
measurement errors of the range $\sigma{D}$, frequency
Doppler shift $\sigma{f}$, and of angles of arrival
of radio waves $\sigma\alpha$ (left - for a magnetically
quiet day, right - for a magnetically disturbed day).
The errors can be evaluated from
the graduated scale, shown in the figure at the right.}
\label{fig3}
\end{figure}

\begin{figure}
\caption
{Schematic representation of the
transionospheric sounding geometry. The axes $z$,
$y$ and $x$ are
directed, respectively, zenithward,
northward $(N)$ and eastward
$(E)$. $P$ -- point of intersection
of Line-of-Sight $(LOS)$ to
the satellite with a horizontal
plane at the height of the
maximum of the ionospheric $F_2$ -- region
$h_{max}$; $S$ --
subionospheric point; and $\Theta_{s}$,
$\alpha_{s}$ -- azimuth
and elevation of the direction  {\bf r}
along LOS to the
satellite.}
\label{fig4}
\end{figure}

\begin{figure}
\caption
{TEC variations $dI(t)$ for the
magnetically quiet day of July 29, 1999
(left, panel $a$) and a
major magnetic storm of July 15, 2000
(right, panel $e$). These
panels also show the names and coordinates
of the $COSA$ and
$BLYT$ stations and GPS satellite numbers
$(PRN)$. Panels
$b$,$c$, $d$, $f$, $g$, $h$ on a logarithmic
scale present the
amplitude spectra $S(F)$ of TEC fluctuations
and derivatives
(left scale in the panels), and the
fluctuation spectra of
$\sigma{D}$, $\sigma{f}$ and $\sigma\alpha$
converted to the
working frequency of 300 $MHz$ (right panel).
For comparison,
spectra from the magnetically quiet day of
July 29, 1999 are
shown in panels $f$, $g$, $h$ by the thin line.
The range of
fluctuation periods is shown in bold type
along the abscissa axis
of panels $d$ and $h$. Panels $b$, $e$ show
the number n of GPS
arrays. The power law character of the
spectra is determined by
the values of slopes $\nu$. The thin black
slant line in panel
$b$ is a fitted straight line of the power
law spectrum.}
\label{fig5}
\end{figure}

\begin{figure}
\caption
{Probability distributions
$P_{\sigma{D}}(\nu)$, $P_{\sigma{f}}(\nu)$ and
$P_{\sigma\alpha}(\nu)$ of the slopes
$\nu_{\sigma{D}}$, $\nu_{\sigma{f}}$ and
$\nu_{\sigma\alpha}$ of the power law spectra of
the errors $\sigma{D}$, $\sigma{f}$ and
$\sigma\alpha$. Mean values of the slopes $\nu$,
and also total numbers of the values of the slopes
of the corresponding spectra use in the analysis
are shown in all panels.}
\label{fig6}
\end{figure}

\begin{figure}
\caption
{Probability distributions
$P_b(\sigma{D})$, $P_b(\sigma{f})$ and
$P_b(\sigma\alpha)$ of the scales
$lg~b_{\sigma{D}}$, $lg~b_{\sigma{f}}$ and
$lg~b_{\sigma\alpha}$ of the power law spectra of
the errors. Mean values of the slopes
$<lg~b_{\sigma{D}}>(lg~b_{\sigma{f}}>,
<lg~b_{\sigma\alpha}>)$ (on a logarithmic scale),
standard deviations, and also total numbers of the
values of the scales of the corresponding spectra
used in the analysis are shown in all panels.}
\label{fig7}
\end{figure}

\begin{thebibliography}{}

\bibitem{}
Afraimovich,~E.~L., Palamartchouk,~K.~S. and
Perevalova,~N.~P.,
GPS radio interferometry of travelling ionospheric disturbances.
{\it J. Atmos. and Solar-Terr. Physics}., V.~60, N.~12,
pp.~1205--1223, 1998.

\bibitem{}
Afraimovich,~E.~L., Kosogorov,~E.~A., Lesyuta,~O.~S.,
Yakovets,~A.~F.,
Ushakov,~I.~I., Geomagnetic control of the spectrum of traveling
ionospheric disturbances based on data from a global GPS network.
{\it Ann. Geophys}., V.~19, N.~7, pp.~723--731, 2001.

\bibitem{}
Afraimovich,~E.~L., Karachenschev,~V.~A., Range, Doppler
frequency and radio wave angle-of-arrival measurement errors for
satellite-based radio engineering systems caused by the Earth's
ionosphere (as deduced using data from the global GPS network),
{\it Proceedings of $VIII$ International Symposium on radar,
navigation and communication}, Voronezh, V.~2, pp.~1405--1416,
2002.

\bibitem{} Davies,~K., Ionospheric Radio Waves.
{\it Blaisdell, Waltham, Mass}., 1969.

\bibitem{} Gajlit,~T.~A., Gusev,~V.~D., Erukhimov,~L.~M.,
Shpiro,~P.~I., Spectrum of the phase fluctuations at the
ionosphere sounding, {\it Izv. Vuzov. Radiofizika}, V.~26, N.~7,
pp.~795--801, 1983.

\bibitem{} Gershman,~B.~N., Erukhimov,~L.~M. and Yashin,~Yu.~Ya.
Wave Phenomena in the Ionosphere and Space Plasma.
{\it Moscow: Nauka}, 1984, p.~392.

\bibitem{} Goodman,~Dj.~M., Aarons,~J., Ionospheric Effects on
Modern Electronic System, {\it Proceedings of the IEEE}, V.~78,
N.~3, pp.~512--528, 1990.

\bibitem{} Hofmann-Wellenhof,~B., Lichtenegger,~H. and
Collins,~J., Global Positioning System: Theory and Practice.
{\it Springer-Verlag, New York}, 1992, p.~327.

\bibitem{} Klobuchar,~J.~A., Ionospheric time-delay algorithm for
single-frequency GPS users, {\it IEEE Transactions of
Aerospace and Electronic System, AES},
V.~23, N.~3, pp.~325--331, 1986.

\bibitem{} Kolosov,~M.~A., Armand,~N.~A. and Yakovlev,~O.~I. The
Propagation of Radio Waves in Space Communication.
{\it Moscow: Svyaz}, 1969, p.~155.

\bibitem{} Komrakov,~G.~P., Skrebkova,~L.~A., Study of parameters
of ionospheric irregularities by the "Interkosmos-Kopernik 500"
satellite, {\it Ionosfernie issledovaniya, Moscow: Sovetskoe
radio}, V.~30, pp.~49--52, 1980.

\bibitem{} Kravtsov,~A.~Yu., Feizulin,~Z.~I. and
Vinogradov,~A.~G. The Propagation of Radio Waves Through the
Earth's Ionosphere. {\it Moscow: Radio i svyaz}, 1983, p.~224.

\bibitem{} Livingston,~R.~C., Rino,~C.~L., McClure,~J.~P.,
Hanson,~W.~B., Spectral characteristics of medium-scale
equatorial F region irregularities, {\it J. Geophys. Res.},
V.~86, pp.~2421--2428, 1981.

\bibitem{} Mannucci,~A.~J., Wilson,~B.~D., Yuan,~D.~N.,
Ho,~C.~M., Lindqwister,~U.~J. and Runge,~T.~F., A global mapping
technique for GPS-drived ionospheric TEC measurements. {\it Radio
Science}, V.~33, pp.~565--582, 1998.

\bibitem{} Schaer,~S., Gurtner,~W. and Feltens,~J., IONEX: The
IONosphere Map EXchange Format Version 1. {\it Proceeding of the
IGS AC Workshop, Darmstadt, Germany, February 9-11; Editor
J.W.Dow}, pp.~233--247, 1998.

\bibitem{} Spoelstra,~T.~A., Kelder,~H., Effects produced by the
ionosphere on radio interferometry. {\it Radio Science}, V.~19,
pp.~779--788, 1984.

\bibitem{} Thompson,~R., Moran,~J. and Swenson,~Dj.
Interferometry and Synthesis in Radio Astronomy. {\it John Wiley
and Sons, Inc.}, 1986.

\bibitem{} Wilson,~B.~D., Mannucci,~A.~J. and Edwards,~C.~D.,
Subdaily northern hemisphere maps using the IGS GPS network. {\it
Radio Science}, V.~30, pp.~639--648, 1995.

\bibitem{} Yakovlev,~O.~I. The Propagation of Radio Waves in
Space. {\it Moscow: Nauka}, 1985, p.~214.

\bibitem{} Yakubov,~V.~P. Doppler Very-Long-Baseline
Interferometry. {\it Tomsk: Vodolei}, 1997, p.~246.

\end{thebibliography}
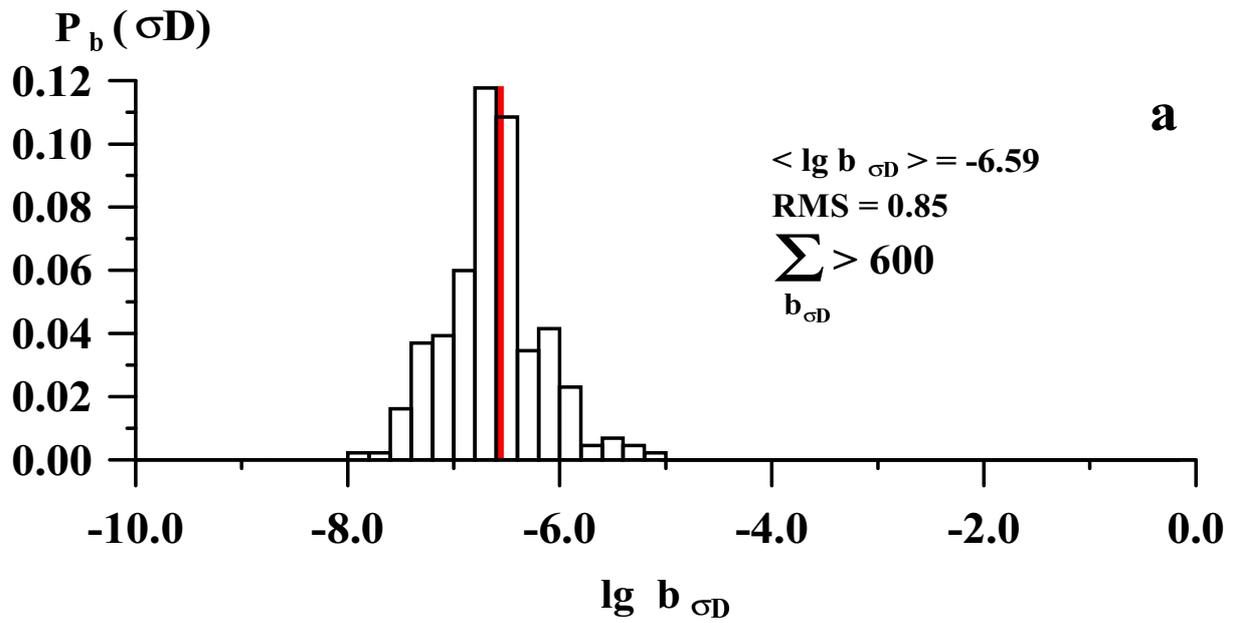
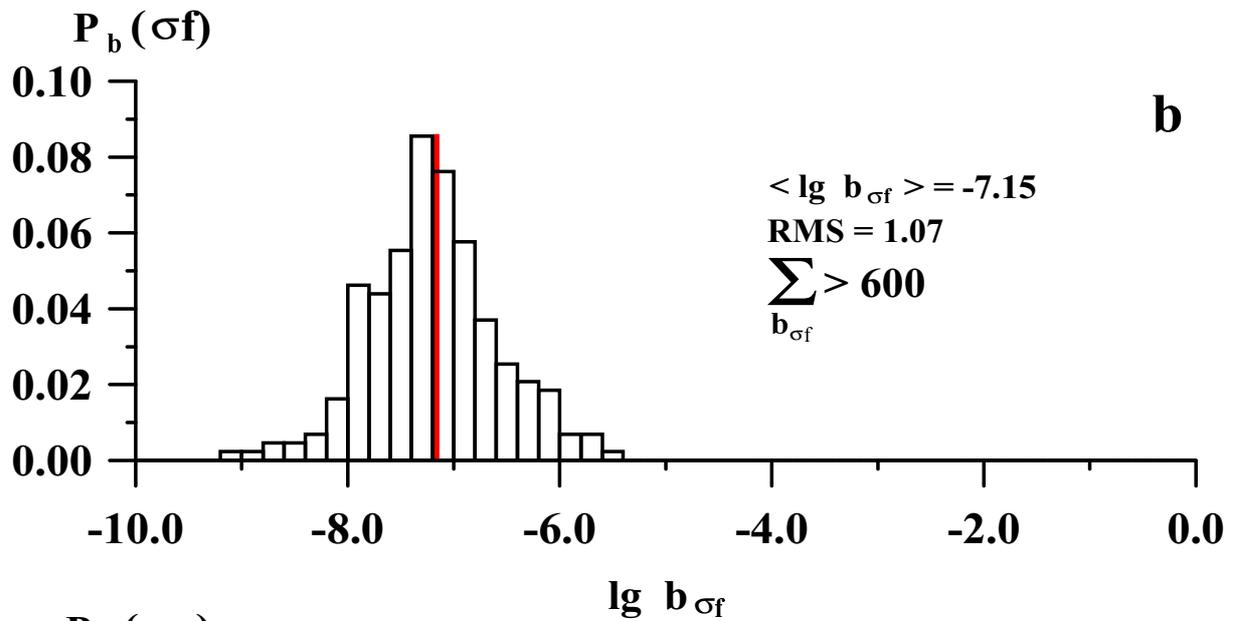
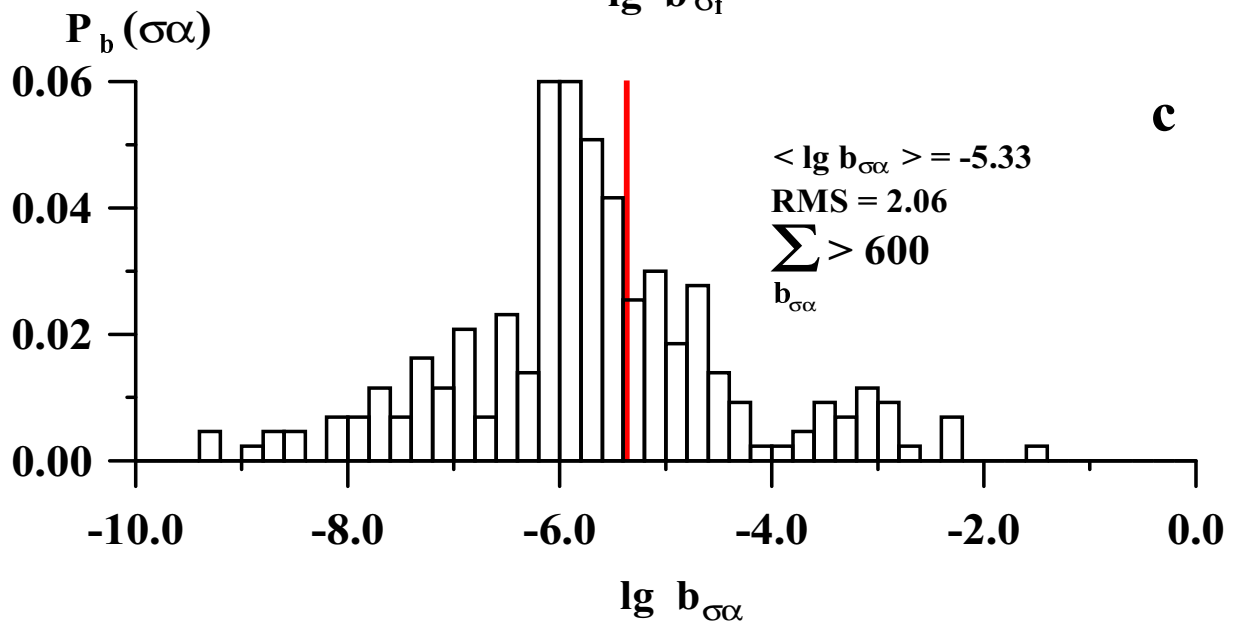
\end{document}